\documentclass[a4paper]{article}
 \usepackage{graphicx}
 \usepackage{amsmath}
 \usepackage{hyperref}
 \usepackage{color}

\begin{document}
 
 \begin{center}
 
 {\bf \Large Computational indeterminism in complex models of social systems  }\\[5mm]

 {\large  Krzysztof Ku{\l}akowski}\\[3mm]

 {\em
 
 Faculty of Physics and Applied Computer Science, AGH University of
 Science and Technology, al. Mickiewicza 30, PL-30059 Krak\'ow,
 Poland\\

 }

 
 {\tt kulakowski@fis.agh.edu.pl}

 \bigskip
 
 \today
 
 \end{center}
 
 \begin{abstract}
We argue that complex models with many parameters do not allow to infer about cause and effect.
\end{abstract}
 
 \noindent
 
 
 \noindent
 

 \bigskip

\section{Classical, discrete, yet unpredictable?}

Applications of computational models to social phenomena met a range of obstacles of different nature and level, from unrepeatability of experiments to lack of interest of social scientists. An objection which is particularly grave states that the concept of determinism - {\it theory that all events, including moral choices, are completely determined by previously existing causes} \cite{1} - fails in social sciences. This objection seems to be grounded on both experimental and moral premises. Namely, our free will is confirmed by our everyday experience and by our moral norms on responsibility for our actions. What, if it appears to be precluded by a deterministic law? In this sense our free will is incompatible with determinism of scientific results on human behaviour. \\

Discussions on freedom of a human being can be traced back to ancient Stoics, but it was not until XIX century when they have been sharpened by the formulation of sociological laws by Auguste Comte. Modern physics added some arguments to this discussion, based on the Heisenberg principle of uncertainty and on the concept of deterministic chaos \cite{2}. However, as it was rightly pointed out by W{\l}adys{\l}aw Tatarkiewicz \cite{3}, indeterminism in physics deals with our inability to determine future states, and not with lack of cause and effect in Nature. Even more, how this indetermism could be overcame by a conscious decision of mind? This question is equivalent to looking for entirely new physics; in fact, the connection of consciousness with quantum mechanics has been explored by Roger Penrose \cite{4}. \\

We note here that the above problem of incompatibility is cured only seemingly by a reference to the law of large numbers, as in \cite{5}; if people decide according to their free will, how the average over these decisions could be determined by a deterministic law? In this case, an accordance of model results with real data should be taken as only fortuitous. The assumption of the free will is, however, hard to justify {\it a priori}. On the other hand, much effort has been done to look for deterministic laws in human behavior \cite{6}; any success, fortuitous or more than that, is perhaps the best thing we can have. For sure, we cannot criticize such approaches until somebody offers a better insight. \\

Both quantum uncertainty and chaotic instabilities are known to produce unpredictable behaviour. Here we are going to move back to third cause of indeterminism, which is present without the former two. The third agent, historically the first, is the influence of an environment. To investigate this influence, we should use a classical (non quantum) model and we should exclude chaotic effects. The latter condition can be met by using the formalism of cellular automata, where time, space and states are discrete \cite{7,8}. We are going to discuss limitations of concepts of cause and effect in a system described by discrete variables.\\

\section{Many causes}

Let us start from an elementary cellular automaton \cite{7} where cells form a linear chain. There are two states of each cell, say 0 and 1. The neighborhood radius is 1; therefore the state of a cell at time $t+1$ depends on the states of this cell and its two neighbours at time $t$. As we know, there is 256 elementary automata; they can be conveniently numbered as the automaton rule written in the decimal system \cite{7}. \\

Let us consider an automaton 11110000. It is equivalent to the rule, where the state of a cell at time $t+1$ is the same as the state of its left neighbour at time $t$. This is explained in detail in Table 1. In this rule, cause and effect can be identified very clearly. The cause is the previous state of the left neighbour. The effect is the actual state of the central cell.\\

\begin{table}
\begin{center}
\begin{tabular}{*{8}{c}}
111&110&101&100&011&010&001&000\\\hline
1&1&1&1&0&0&0&0\\\hline
\end{tabular}
\end{center}
\caption{The rule of automaton 11110000 = 240. In the upper line, the list is given of all possible states of three cells. In the bottom line, the resulting state of the central cell is given under each state  of three cells. The bottom line can be read as a number 11110000, what is equal to 240 when translated from the binary to the decimal system. This particular rule means that the central cell at time $t+1$ is equal to the state of its left neighbour at time $t$.}
\label{table1}
\end{table}

To give an example of this rule, let us assume that some group of gentlemen established a political club with an old-style sitting room. There, they are sitting around a table in always the same order: A has Z at left hand and B at right, B has A at left and C at right and so on. The rule is that at each national presidential elections, each club member votes in the same way as his left neighbour voted in the preceding elections: for Democrats or Republicans.\\

Now let us consider another rule 10010110 = 150, shown in Table 2. In this case,  a change of state of each cell at time $t$ leads to a change of state of the central cell at time $t+1$. There is no single cell which could be identified as a cause of the central cell. The state of the central cell depends on the previous state of all three cells. This rule is crucial for our argumentation here. Obviously, there are many causes which lead to the effect. Keeping our previous example of the political club, it would be not easy for a club member to answer the question: why did you vote to Democrats?\\

\begin{table}
\begin{center}
\begin{tabular}{*{8}{c}}
111&110&101&100&011&010&001&000\\\hline
1&0&0&1&0&1&1&0\\\hline
\end{tabular}
\end{center}
\caption{The rule of automaton 10010110 = 150. The rule is designed in the way that a change of state of each cell at time $t$ leads to a change of state of the central cell at time $t+1$.}
\label{table2}
\end{table}

Yet, our rule 150 is simple enough to find another cause than the state of one cell. Namely, it is the number of zeros in the row of three cells at time $t$. If this number is even, the outcome is zero; otherwise it is 1. However, when the neighborhood radius is as large as the system size, the problem to find a simple cause for a given rule has no general solution; in most cases, there are multiple causes.\\

To visualise the difficulty, we can imagine a large club with a hundred or more members sitting together. Suppose that each person has his own rule to determine how to vote during next elections. Each personal rule can be different and its outcome can depend of a different and large subset of club members. For simplicity, we can assume that the rules do not change in time. Still, the task how to find the rules from the complete data on individual voting is computationally hard and it cannot be accomplished in a reasonable time. This means that we cannot predict the outcome of a next election on the basis of observation. The problem is deterministic {\it per se}, but the rule remains unknown. Even the deterministic character of the problem can be hard to verify. This is because the number of states increases exponentially with the system size, represented here by the number of club members. The criterion of the deterministic character could be if two identical states a times $t$ and $t'$ produce two identical outcomes at $t+1$ and $t'+1$. The devil is in that the time of experiment should be long enough to observe such coincidence of states: in general, the difference $t'-t$ also increases exponentially with the system size.\\

\section{Conclusions on modeling}

Suppose that we intend to build a realistic model of a social system, with the purpose to reproduce observed facts. On the contrary to a mathematical pendulum, a social system cannot be introduced to a computer as a whole; some amount of reductionism is always necessary. This means that we never know everything on a social system. Still, we are often tempted to capture more and more details. The problem is not to lose control of our model. The purpose of modeling is to obtain relations 'cause-effect'. In social systems, presumably an effect has multiple causes. A given value of each parameter can be a cause. Then, unless the values of the model parameters are either taken directly from other experiments, they should be varied to check the result. However, if the number of parameters is too large, this procedure cannot be accomplished. This means that in complex models the concepts of cause and effect do not work. In this way, determinism fails for purely computational reasons.\\

This argument allows to state that complex models with many parameters are not useful. Even if an accordance of the model result with reality is obtained, it is likely that one can obtain different results, including those contradict to reality, when the parameters are changed. If this is not the case, some parameters are irrelevant and they should be removed from the model. With many parameters to handle, even the latter possibility cannot be verified.\\

It is worthwhile to note that the above argumentation is by no means a support for the concept of free will. The relation of social scientists to this concept is twofold. As human beings, we believe that free will is indispensable for moral reasons; as a cultural achievement of human race it is a social good of the highest value. As scientists, we doubt if we can prove its existence and we are deeply interested in identification of situations where free will is limited. Above we argued that if in our decisions we are just complex deterministic machines, it is likely that we cannot recognize this. \\


\begin{thebibliography}{99}
\bibitem{8} Automata where information can spread without bariers are termed 'chaotic' by analogy to the deterministic chaos. As we know, the maximal velocity of information in a cellular automaton is equal to the radius of neighborhood by one time step. Therefore a change of initial conditions (which cannot be arbitrarily small because of discreteness of variables) can spread at most linearly with time. On the contrary, the essence of the deterministic chaos is a sensitivity to an arbitrarily small variation of initial conditions, which spreads exponentially with time.
\end{thebibliography}
 \end{document}